\documentclass[twocolumn,amsmath,amssymb,showpacs,prl]{revtex4-1}
\usepackage[latin1]{inputenc}

\begin{document}

\title{Semiclassical Strings in \(AdS_5\times S^5\) and Automorphic Functions\\}
\author{Michael Pawellek}
\affiliation{Department of Theoretical Physics, \\
 Royal Institute of Technology (KTH)- AlbaNova University Center, \\
 Roslagstullbacken 21, 106 91 Stockholm, Sweden}
\email{pawellek@kth.se}

\begin{abstract}
 Using AdS/CFT we derive from the folded spinning string ordinary differential equations for the anomalous dimension of the dual \(\mathcal{N}=4\) SYM twist-two operators at strong coupling. We show that for large spin the asymptotic solutions have the Gribov-Lipatov reciprocity 
 property. To obtain this result we use a hidden modular invariance of the energy-spin relation of the folded spinning string. 
 Further we identify the Moch-Vermaseren-Vogt (MVV) relations, which were first recognized in plain QCD calculations, as the recurrence 
 relations of the asymptotic series ansatz.   
\end{abstract}
\maketitle
 
\section{Introduction}
Since the early days of QCD \cite{Gros} it is known that the anomalous dimensions \(\gamma(S)\) of twist operators 
 \(\mathrm{Tr}(\Phi D^S\Phi)\) are related to parton splitting functions, which gives the probability of partons to split into another parton with certain momentum fractions. Nevertheless it is a recent observation that the 
coefficient functions at three loop order in perturbation expansion of \(\gamma(S)\) are already determined by the two loop results 
\cite{Moch, Moch2, Doks}.
In a large \(S\) expansion the terms organize as \cite{Bass, Becc}
\begin{equation}
 \gamma(S)=f\ln(S)+f_c+f_{11}\frac{\ln(S)}{S}+f_{10}\frac{1}{S}+\mathcal{O}(S^{-2}),
\end{equation}
where 
\begin{equation}
 f_{11}=\frac{1}{2}f^2, \qquad f_{10}=\frac{1}{2}f\,f_c
\end{equation}
are the MVV relations. 
In \cite{Bass, Becc} it was argued that similar relations hold also for coefficient functions of the higher order terms with odd
powers in \(S\) 
provided the anomalous dimension has the parity preserving asymptotic expansion in the conformal spin 
\(s=S+\frac{1}{2}\gamma(S)\):
\begin{equation}\label{eq:parity}
 \gamma(S)=\mathrm{f}(S+\frac{1}{2}\gamma)\to g\ln(s)+g_c+\frac{g_1}{s^2}+\frac{g_2}{s^4}+\mathcal{O}(s^{-6}),
\end{equation}
where \(g_i\) are polynomials in \(\ln(s)\).
The parity preserving property implies also the Gribov-Lipatov reciprocity of the splitting functions \cite{Grib, Bass}.

The conjectured AdS/CFT correspondence \cite{Mald, Gubs, Gub2} relates 
the anomalous dimension of twist operators in \(\mathcal{N}=4\) \(SU(N)\) SYM at strong coupling to the energy \(E(S)\) of semiclassical spinning string configurations with spin \(S\) in \(AdS_5\times S^5\). For recent reviews of the anomalous dimension
and Gribov-Lipatov reciprocity of twist operators in context of AdS/CFT see \cite{Bec3}.

In this letter we will use AdS/CFT to show that at strong coupling the MVV relations and the parity preserving and reciprocity properties can be derived from asymptotic series solutions of nonlinear ordinary differential equations. We obtain these equations by using 
the close relation of the energy-spin relation of the dual spinning string to automorphic functions and Picard-Fuchs equations for complete elliptic integrals.

\section{The folded spinning string}
In the subspace \(AdS_3\) of \(AdS_5\times S^5\) with metric  
\begin{equation}
 \mathrm{d}s^2=-\cosh^2\rho\mathrm{d}t^2+\mathrm{d}\rho^2+\sinh^2\rho\mathrm{d}\phi^2,
\end{equation}
the folded spinning string can be described by the ansatz \cite{Gub2, Frol} 
\begin{equation}\label{eq:RotAnsatz}
 t=\kappa\tau,\;\;\phi=\omega\tau,\;\;  \rho(\sigma)=\rho(\sigma+2\pi).
\end{equation}
The explicit solution of the equations of motions is given in terms of Jacobi's elliptic functions
\begin{equation}
 \cosh^2\rho(\sigma)=\frac{1}{1-k^2}\mathrm{dn}^2(\omega\sigma+\mathbb{K}|k^2),
\end{equation}
and from the periodicity constraints one gets relations between the parameters in the ansatz:
\begin{equation}
 \kappa=\frac{2}{\pi}k\mathbb{K},\qquad \omega=\frac{2}{\pi}\mathbb{K}.
\end{equation}

Also the energy and the spin can be parameterized in terms of complete elliptic integrals
\begin{eqnarray}
 E&=&\sqrt{\lambda}\frac{2}{\pi}\frac{k}{1-k^2}\mathbb{E},\\
 S&=&\sqrt{\lambda}\frac{2}{\pi}\left[\frac{1}{1-k^2}\mathbb{E}-\mathbb{K}\right],
\end{eqnarray}
where the modulus parameter \(k\) is related to the string length \(\rho_0\) as
\begin{equation}
 \tanh\rho_0=k.
\end{equation} 
In the semiclassical limit 
\begin{equation}
 \lambda,S,E\to \infty,\qquad \text{with }\, \mathcal{E}=\frac{E}{\sqrt{\lambda}},\qquad \mathcal{S}=\frac{S}{\sqrt{\lambda}}\; \text{ fix}
\end{equation}
the energy can be expanded as
\begin{equation}
 E=\sqrt{\lambda}\mathcal{E}+E_1+\mathcal{O}\left(1/\sqrt{\lambda}\right),
\end{equation}
where the classical contribution for \(\mathcal{S}\to\infty\) is given as (with \(\mathcal{\bar S}=8\pi\mathcal{S}\)) \cite{Becc}:
 \begin{eqnarray}\label{eq:classicalenergy}
& & \mathcal{E}(\mathcal{S})=\mathcal{S}+\frac{1}{\pi}\left(\ln\mathcal{\bar S}-1\right)+\frac{1}{2\pi^2\mathcal{S}}
 \left(\ln\mathcal{\bar S}-1\right)-\nonumber\\
& & -\frac{1}{16\pi^3\mathcal S^2}\left(2\ln^2\mathcal{\bar S}-9\ln\mathcal{\bar S}+5\right)+\\
& &+\frac{1}{48\pi^4\mathcal{S}^3}\left(2\ln^3\mathcal{\bar S}-18\ln^2\mathcal{\bar S}+33\ln\mathcal{\bar S}-14\right)+
 \mathcal{O}(\mathcal{S}^{-4}).\nonumber
\end{eqnarray}
In principle one has to insert the inverse function of \(\mathcal{S}\) into \(\mathcal{E}\) to get the expression for \(\mathcal{E}(\mathcal{S})\).
Since this was not possible, so far only this series expression was obtained.

\section{Picard-Fuchs differential equations of elliptic integrals}
In order to understand the modular derivatives of \(\mathcal{E}\) and \(\mathcal{S}\) we will mention some classic
results about Picard-Fuchs equations of complete elliptic integrals. Making the connection to the corresponding
Abelian periods it will then be straightforward to generalize this setting to the general hyperelliptic case \cite{Fuch, Koen}.

The classic canonical form of the complete elliptic integrals are defined by
\begin{eqnarray}
 \mathbb{K}&=&\int_0^1\frac{\mathrm{d}x}{\sqrt{(1-x^2)(1-k^2x^2)}},\nonumber\\
 \mathbb{E}&=&\int_0^1\frac{\mathrm{d}x\sqrt{1-k^2x^2}}{\sqrt{1-x^2}}.
\end{eqnarray}
In algebraic geometry one considers instead the Abelian periods
\begin{equation}\label{eq:abelperiod}
 I_1=\int_0^1\frac{\mathrm{d}z}{\sqrt{P(z)}}\qquad I_2=\int_0^1\frac{\mathrm{d}z\,z}{\sqrt{P(z)}},
\end{equation}
with \( P(z)=z(1-z)(1-k^2z)\). Both expressions are related by (with \(z=x^2\))
\begin{equation}\label{eq:elliptic}
 \mathbb{K}=\frac{1}{2}I_1,\qquad \mathbb{E}=\frac{1}{2}I_1-\frac{k^2}{2}I_2.
\end{equation}
Therefore the complete elliptic integrals are related to the Abelian periods by an additive combination.

Although traditionally the dependence on the elliptic modulus \(k\) is not mentioned explicitly one should keep in mind that
the Abelian periods and complete elliptic integrals are still functions of \(k\). It is a nice property of the Abelian periods that
their derivatives with respect to \(k\) can again be expressed as an additive combinations of the original Abelian periods 
(with \(k'^2=1-k^2\)):
\begin{equation}
 \frac{\mathrm{d}I_1}{\mathrm{d}k}=\frac{k}{k'^2}(I_1-I_2),\;\;\frac{\mathrm{d}I_2}{\mathrm{d}k}=\frac{1}{kk'^2}(I_1-(2-k^2)I_2).
\end{equation}
We avoid the term 'linear combination' since the coefficients in front of the Abelian periods are not independent constants but are again
expressed in terms of the elliptic modulus \(k\).

Using now (\ref{eq:elliptic}) one can immediately write down the modular derivatives for the complete elliptic integral:
\begin{equation}\label{eq:modulderiv}
 \frac{\mathrm{d}\mathbb{K}}{\mathrm{d}k}=\frac{1}{kk'^2}(\mathbb{E}-k'^2\mathbb{K}),\qquad 
 \frac{\mathrm{d}\mathbb{E}}{\mathrm{d}k}=\frac{1}{k}(\mathbb{E}-\mathbb{K}).
\end{equation}
Repeating this procedure with the second derivative one finds the classic result by Legendre that the complete elliptic integral satisfies a second order linear differential equation
\begin{equation}\label{eq:moduldiff}
 kk'^2\frac{\mathrm{d}^2\mathbb{K}}{\mathrm{d}k^2}+(1-3k^2)\frac{\mathrm{d}\mathbb{K}}{\mathrm{d}k}-k\mathbb{K}=0.
\end{equation}
This equation is today understood as a special case of Picard-Fuchs equations.
By the substitution \(x=k^2\) the differential equation (\ref{eq:moduldiff}) transforms to a hypergeometric equation,
which gives the well known expression for the complete elliptic integral in terms of hypergeometric functions
\begin{equation}
 \mathbb{K}=\frac{\pi}{2}{}_2F_1\left(\frac{1}{2},\frac{1}{2},1;k^2\right).
\end{equation}

\section{The energy-spin function}
We can now take a look at the energy and spin of the folded string from the view point of Abelian periods and their
modular derivatives:
\begin{equation}\label{eq:Cartancharge}
 \mathcal{E}=\frac{2}{\pi}\frac{k}{1-k^2}\mathbb{E},\qquad
 \mathcal{S}=\frac{2}{\pi}\left[\frac{1}{1-k^2}\mathbb{E}-\mathbb{K}\right].
\end{equation}
Since (\ref{eq:Cartancharge}) are additive combinations of the complete elliptic integrals, they are also additive combinations of the Abelian periods.
And therefore they inherit the property that their modular derivatives can again be expressed not only as additive combination of complete
elliptic integrals or Abelian periods but also in terms of \(\mathcal{E}\) and \(\mathcal{S}\) again:
\begin{equation}\label{eq:Cartanderiv}
 \frac{\mathrm{d}\mathcal{E}}{\mathrm{d}k}=\frac{1}{1-k^2}\left[\mathcal{S}+\frac{1}{k}\mathcal{E}\right],\;\;\;
 \frac{\mathrm{d}\mathcal{S}}{\mathrm{d}k}=\frac{1}{1-k^2}\left[k\mathcal{S}+\mathcal{E}\right].
\end{equation}
This is not really a surprise but the essential point is: Although we have no analytic expression for \(\mathcal{E}(\mathcal{S})\) we can use now the results (\ref{eq:Cartanderiv})
to obtain a very nice compact expression for its derivative: 
\begin{equation}\label{eq:firstderiv}
 \frac{\mathrm{d}\mathcal{E}}{\mathrm{d}\mathcal{S}}=\frac{1}{k},
\end{equation}
where the elliptic modulus \(k\) should be understood as function of \(\mathcal{S}\).

On the other hand the elliptic modulus has a representation in terms of theta constants
\begin{equation}
 k=\frac{\vartheta_{10}^2(0,\tau)}{\vartheta_{00}^2(0,\tau)},
\end{equation}
which has the property that it is invariant under modular transformations \(\Gamma(2)\) of \(\tau\)\cite{Mumf}, it is an
automorphic function. Further, transformations under the full modular group \(\Gamma\)  may be
related to recently observed duality properties of the spinning string \cite{Geor}.

\(\mathcal{E}\) and \(\mathcal{S}\) are a very special combination of the complete elliptic integrals,
since from (\ref{eq:modulderiv}) one can see that e.g. \(\mathrm{d}\mathbb{K}/\mathrm{d}\mathbb{E}\) is not an algebraic expression
of an automorphic function.

We can proceed further and consider the second derivative, given by
\begin{eqnarray}
 \frac{\mathrm{d}^2\mathcal{E}}{\mathrm{d}\mathcal{S}^2}&=&\frac{\mathrm{d}}{\mathrm{d}\mathcal{S}}
 \left(\frac{\mathrm{d}\mathcal{E}}{\mathrm{d}\mathcal{S}}\right)=
 \left(\frac{\mathrm{d}\mathcal{S}}{\mathrm{d}k}\right)^{-1}\frac{\mathrm{d}}{\mathrm{d}k}\left(\frac{1}{k}\right)=\nonumber\\
 &=&-\frac{1-k^2}{k^2}\frac{1}{k\mathcal{S}+\mathcal{E}}.
\end{eqnarray}
Using now (\ref{eq:firstderiv}) for \(k\), we obtain a nonlinear second order ordinary differential equation of third degree:
\begin{equation}\label{eq:nonlin}
 \left(\mathcal{S}+\mathcal{E}(\mathcal{S})\frac{\mathrm{d}\mathcal{E}}{\mathrm{d}\mathcal{S}}\right)
 \frac{\mathrm{d}^2\mathcal{E}}{\mathrm{d}\mathcal{S}^2}+\left(\frac{\mathrm{d}\mathcal{E}}{\mathrm{d}\mathcal{S}}\right)^3-
 \frac{\mathrm{d}\mathcal{E}}{\mathrm{d}\mathcal{S}}=0.
\end{equation}
This is the main result of this section. The energy-spin function \(\mathcal{E}(\mathcal{S})\) of the folded string is a solution of (\ref{eq:nonlin}).

Motivated by the previous result (\ref{eq:classicalenergy}) one can use the following asymptotic ansatz for 
\(\mathcal{S}\to\infty\) as a solution of (\ref{eq:nonlin}):
\begin{eqnarray}
 \mathcal{E}(\mathcal{S})&=&\mathcal{S}+f\ln\mathcal{S}+f_c+\frac{f_{11}\ln\mathcal{S}+f_{10}}{\mathcal{S}}+\nonumber\\
 & &+\frac{f_{22}\ln^2\mathcal{S}+f_{21}\ln\mathcal{S}+f_{20}}{\mathcal{S}^2}+\nonumber\\
 & &+\frac{f_{33}\ln^3\mathcal{S}+f_{32}\ln^2\mathcal{S}+f_{31}\ln\mathcal{S}+f_{30}}{\mathcal{S}^3}.
\end{eqnarray}
Then one obtains recurrence relations between the coefficients
\begin{eqnarray}\label{eq:recurrence1}
 f_{11}&=&\frac{1}{2}f^2,\qquad f_{10}=\frac{1}{2}ff_c,\qquad f_{22}=-\frac{1}{8}f^3, \nonumber\\
 f_{21}&=&\frac{1}{16}f^2(5f-4f_c), \nonumber\\
 f_{20}&=&\frac{1}{16}f(2f^2+5ff_c-2f_c^2),\nonumber\\
 f_{33}&=&\frac{1}{24}f^4,\qquad f_{32}=\frac{1}{8}f^3(-2f+f_c),\nonumber\\
 f_{31}&=&\frac{1}{16}f^2(f^2-8ff_c+2f_c^2),\nonumber\\
 f_{30}&=&\frac{1}{48}f(3f^3+3f^2f_c-12ff_c^2+2f_c^3).
\end{eqnarray}
Since we have a second order differential equation 
all coefficients are determined when \(f\) and \(f_c\) are fixed by 
some initial conditions. Apparently these are the MVV relations previously derived by assuming the parity preserving property 
(\ref{eq:parity}). We see now that only two free coefficients are enough to determine the solution.

As a check one can convince oneself that the coefficients of the classical energy (\ref{eq:classicalenergy}) indeed satisfy the relations 
(\ref{eq:recurrence1}).

\subsection{The weak involution property}
 For the function \(\mathcal{S}(\mathcal{E})\), which is the inverse of \(\mathcal{E}(\mathcal{S})\), we have
 \begin{equation}
 \frac{\mathrm{d}\mathcal{S}}{\mathrm{d}\mathcal{E}}=k,
 \end{equation}
from which we can derive the corresponding differential equation
\begin{equation}
 \left(\mathcal{E}+\mathcal{S}(\mathcal{E})\frac{\mathrm{d}\mathcal{S}}{\mathrm{d}\mathcal{E}}\right)
 \frac{\mathrm{d}^2\mathcal{S}}{\mathrm{d}\mathcal{E}^2}+
 \left(\frac{\mathrm{d}\mathcal{S}}{\mathrm{d}\mathcal{E}}\right)^3-\frac{\mathrm{d}\mathcal{S}}{\mathrm{d}\mathcal{E}}=0
\end{equation}
Comparing with (\ref{eq:nonlin}) we see that the equations are invariant under \(\mathcal{S}\to\mathcal{E}\) and \(\mathcal{E}\to\mathcal{S}\).
This means that a function \(f_1(x)\) and also its inverse \(f_2(x)=f_1^{-1}(x)\) are solutions of the same  differential equation: 
\begin{equation}
 f_1(\mathcal{E})=\mathcal{S},\qquad f_2(\mathcal{S})=\mathcal{E},
\end{equation}
%which means that \(f_1(x)\) and its inverse \(f_2(x)=f_1^{-1}(x)\) are solutions of the same differential equation
%which is also a consequence of
%\begin{equation}
% \frac{\mathrm{d}f_2}{\mathrm{d}\mathcal{S}}\frac{\mathrm{d}f_1}{\mathrm{d}\mathcal{E}}=\frac{1}{k}k=1.
%\end{equation}

\subsection{The reciprocity property}
If the anomalous dimension  \(\gamma=\mathcal{E}-\mathcal{S}\) satisfies the parity preserving relation (\ref{eq:parity}) it should be 
possible to write it as a function of the generalized spin variable \(s_0=\mathcal{E}+\mathcal{S}\) \cite{Becc}.

It is now easy to check that also the derivative of this function has the modular invariance property: 
\begin{equation}
 \frac{\mathrm{d}(\mathcal{E}-\mathcal{S})}{\mathrm{d}(\mathcal{E}+\mathcal{S})}=\frac{1-k}{1+k}.
\end{equation}
Along the same line as for \(\mathcal{E}(\mathcal{S})\) we find the differential equation for \(\gamma(s_0)\) as
\begin{equation}
 \left(s_0+\gamma(s_0)\frac{\mathrm{d}\gamma}{\mathrm{d}s_0}\right)\frac{\mathrm{d}^2\gamma}{\mathrm{d}s_0^2}-
 \left(\frac{\mathrm{d}\gamma}{\mathrm{d}s_0}\right)^3+\frac{\mathrm{d}\gamma}{\mathrm{d}s_0}=0.
\end{equation}
One can make a similar series ansatz for \(\gamma(s_0)\) as for \(\mathcal{E}(\mathcal{S})\):
\begin{eqnarray}
 \gamma(s_0)=g_0s_0+g\ln(s_0)+g_c+\sum_{n=1}\frac{a_n(\ln(s_0))}{s_0^{n}},\nonumber\\
  a_n(\ln(s_0))=\sum_{m=0}g_{nm}\ln^m(s_0).
\end{eqnarray}
Now we find the following recurrence relations:
\begin{eqnarray}
 g_0&=&0,\qquad g_{10}=g_{11}=0,\qquad g_{22}=0,\nonumber\\
 g_{21}&=&\frac{1}{4}g^3,\qquad g_{20}=\frac{1}{4}g^2(2g+g_c),\nonumber\\
 g_{33}&=&g_{32}=g_{31}=g_{30}=0,\qquad g_{44}=g_{43}=0,\nonumber\\
 g_{42}&=&-\frac{1}{8}g^5,\qquad
 g_{41}=-\frac{1}{64}g^4(23g+16g_c)\nonumber\\
 g_{40}&=&-\frac{1}{128}g^3(35g^2+46gg_c+16g_c^2),\nonumber\\
 g_{55}&=&g_{54}=g_{53}=g_{52}=g_{51}=g_{50}=0.
\end{eqnarray}
The function \(\gamma(s_0)\) has all the desired properties \cite{Becc,Bec3}
\begin{itemize}
 \item Reciprocity: Only even negative powers of \(s_0\) appear in the expansion,
 \item The highest order of \(\ln^n(s_0)\) does not appear in the expansion.
\end{itemize} 

\subsection{Short string expansion}
 It is also easy to obtain from (\ref{eq:nonlin}) the short string expansion \(\mathcal{S}\to 0\) to arbitrary high order:
\begin{equation}\label{eq:shortexp}
 \mathcal{E}=\sqrt{2\mathcal{S}}\sum_{n=0}a_n\mathcal{S}^n,
\end{equation}
with the following coefficients
\begin{eqnarray}
 a_1&=&\frac{3}{8}\frac{1}{a_0},\qquad a_2=-\frac{21}{128}\frac{1}{a_0^3},\qquad a_3=\frac{187}{1024}\frac{1}{a_0^5},\nonumber\\
 a_4&=&-\frac{9261}{32768}\frac{1}{a_0^7},\qquad a_5=\frac{136245}{262144}\frac{1}{a_0^9}.
\end{eqnarray}
This confirms and extends previous results for \(a_0=1\)\cite{Tirz}. A further numerical comparison of the asymptotic series (\ref{eq:shortexp}) 
with the exact result (\ref{eq:Cartancharge}) shows good agreement for \(\mathcal{S}<0.4\).
\section{Discussion}
We have shown that the energy-spin relation of the semiclassical folded spinning string has to satisfies a nonlinear second order
differential equation. Using the AdS/CFT correspondence the recurrence relations for the coefficients of asymptotic series ansatz at \(\mathcal{S}\to\infty\) are identified as the MVV relations for the anomalous dimensions of the twist operators in the dual gauge theory. 

The analog asymptotic solution of the differential equations for the anomalous dimension expressed in terms of the conformal spin 
has the reciprocity property, an expansion only in even negative powers of \(S\). So far this property was an 
ad hoc assumption in order
to derive the MVV relations. Now we see it is inherit from the modular properties of the spinning string configuration.
Since the corresponding equations are of second order, there remain two undetermined coefficients, the scaling or cusp functions \(f\) and
the virtual scaling function \(f_c\) \cite{Frey}. Although the differential equation and the recurrence relations are derived from the classical string one can check by using the results of \cite{Bec2} that the relations are also satisfied up to \(\mathcal{O}(S^{-2})\) including the 1-loop
quantum corrections. 

In order to generalize our results to other spin configurations we like to point out that it is not necessary to have a closed expression for
the constants of motion. All one needs are the integral expressions in terms of Abelian periods as (\ref{eq:abelperiod}) in the elliptic case. Integral expressions are accessible for a broad range of semiclassical string configuration \cite{Arut}. From these in principle the corresponding Picard-Fuchs equations can be derived in order to find some automorphic invariants. 

It would also be interesting to understand the relation of the nonlinear differential equations to certain integral equations of the anomalous dimension, which follow from the asymptotic Bethe ansatz \cite{Beis, Bas2, Fior}.

I thank T. M\aa nsson and A. Tseytlin for useful comments on the manuscript. This work was supported by the G\"oran Gustafsson Foundation.

\end{document}